\documentclass[10pt,conference]{IEEEtran}
\usepackage{epsfig}
\usepackage{color,subfigure}
\usepackage{amsthm, amssymb, mathrsfs}
\usepackage{amsmath}
\usepackage{graphicx}
\usepackage{array}
\psfull
\DeclareMathOperator{\tr}{tr}

\def\gamth{\gamma_{\text{th}}}

\def \Q{{\mathbf{Q}}}
\def \X{{\mathbf{X}}}
\def \I{{\mathbf{I}}}
\def \H{{\mathbf{H}}}
\def \P{{\mathbf{P}}}
\def \CN{{\mathcal{CN}}}
\def \Pfa{{P_{\text{fa}}}}
\def \C{{\mathcal{C}}}

\def \Rcap{{\hat{\mathbf{R}}}}
\begin{document}
\title{Blind Non-parametric Statistics for Multichannel Detection Based on Statistical Covariances}
\author{\IEEEauthorblockN{Vidyadhar Upadhya and~Devendra Jalihal}
\IEEEauthorblockA{Department of Electrical Engineering\\
Indian Institute of Technology Madras, India\\
Email: ee08s011, dj@ee.iitm.ac.in}
}
\maketitle
\begin{abstract}

We consider the problem of detecting the presence of a spatially correlated multichannel signal corrupted by additive Gaussian noise (i.i.d across sensors). No prior knowledge is assumed about the system parameters such as the noise variance, number of sources and correlation among signals. It is well known that the GLRT statistics for this composite hypothesis testing problem are  asymptotically optimal and sensitive to variation in system model or its parameter. To address these shortcomings we present a few non-parametric statistics which are functions of the elements of Bartlett decomposed sample covariance matrix. They are designed such that the detection performance is immune to the uncertainty in the knowledge of noise variance. The analysis presented verifies the invariability of threshold value and identifies a few specific scenarios where the proposed  statistics have better performance compared to GLRT  statistics. The sensitivity of the statistic to correlation among streams, number of sources and sample size at low signal to noise ratio are discussed.
\end{abstract}
\section{Introduction}
Detecting the presence of signal affected by channel impairments and corrupted by additive noise is encountered in a variety of array processing applications. The goal is to classify the observation into one of the two possibilities \cite{leshem}, i.e., signal present/not present. The detection statistic for this binary hypothesis testing problem is said to be optimal in the Neyman-Pearson (N-P) sense if it maximizes detection probability ($P_d$) for a fixed false alarm level ($P_{fa}$). It is well known that the likelihood ratio test (LRT) assures optimality in the N-P sense if the two distributions are known precisely. 
However, in the case of system with uncertain parameters the hypotheses are composite. In such case, the statistic which is optimal in the N-P sense  irrespective of the value taken by the uncertain parameters
is called uniformly most powerful statistic. There is no straightforward procedure to find it and they don't exist in many real life scenarios \cite{det_steven}. An alternate approach is to obtain the asymptotically optimal detection statistic called the generalized likelihood ratio test (GLRT) by replacing the uncertain parameters with their maximum likelihood estimate (MLE) in the likelihood ratio. 
 
The MLE of the covariances under the two hypotheses for the multichannel signal corrupted by additive Gaussian noise observation model is the function of sample covariance matrix (SCM), because it is the sufficient statistic \cite{mulvar_TWAnderson}. A few variations in the final form of the GLRT statistic depending on the prior information or assumption about the system model and its parameters are discussed in the following paragraph.

When the noise across sensors are assumed independent resulting in diagonal covariance structure GLRT reduces to coherence ratio test \cite{leshem} and under i.i.d assumption resulting in homogeneous diagonal covariance structure GLRT reduces to sphericity test \cite{TSph_Mauchly}.
In addition, the information about the number of sources ($N_t$) contributes in improving the detection strategy by estimating the noise from the latent roots of the model. This method is advantageous when $N_t$ is less than the number of sensors, $N_r$. Assuming the exact knowledge of $N_t$, the GLRT reduces to \textit{reduced sphericity test} \cite{rankP_ramirez} which is interpreted as a measure of sphericity of the sample covariance space to the noise subspace. If there is only a single source and noise variance ($\sigma^2$) is assumed to be known, GLRT reduces to 
maximum eigenvalue test, also known as Roy's largest root test (RLRT) \cite{mulAnt_abbas}. It can also be used as a non-parametric statistic when number of sources is more than one.

The GLRT statistics are known to be sensitive to the prior information or assumption about the system model and its parameters. Also, in practical scenarios no information regarding the data will be available at the  
detector and the sample size will also be limited. To address the shortcomings with these techniques we resort to non-parametric statistics which exploit the spatial correlation across sensors similar to the ones proposed in \cite{robust_zeng} and references therein. The covariance absolute value (CAV) statistic, proposed in \cite{stat_cov} belongs to this category which operates directly on the elements of SCM. It was proposed for the real system model and the test statistic was defined as,
\begin{equation}
 T(\Rcap)={\sum\limits_{i=1}^{N_r} \sum\limits_{j=1}^{N_r} | r_{ij} |}\left/{\sum\limits_{i=1}^{N_r}  r_{ii} }\right.  \label{eq_CAV}
\end{equation}
where $r_{ij}$ represents the $(i,j)^{th}$ element of SCM, $\Rcap$. 

The CAV statistic is used as an \emph{ad hoc} measure to identify the contribution of off-diagonal elements. Under the null hypothesis, the SCM is diagonal due to spatially uncorrelated noise. Hence the CAV statistic approaches unity under the null hypothesis and is greater than unity under the alternate hypothesis due to the existence of  correlation either in the signalling method or when induced spatially. Due to the effectiveness of CAV its performance is used as a reference to compare the blind statistics with the GLRT statistics \cite{tugnait2012multiple,rankP_ramirez,jin2012performance}. This motivates us to look into other forms of covariance
based ratios which can outperform the well established CAV.

\noindent Our previous work \cite{NCC_vidya} extends the analysis of CAV to include the complex data considering its equivalent form as: 
\begin{equation}
 T(\Rcap)={\sum\limits_{1\le i<j\le N_r}  | r_{ij} |}\left/{\sum\limits_{i=1}^{N_r}  r_{ii} }\right.  \label{mod_CAV}
\end{equation}
Due to the dependent nature of the numerator and the denominator terms, the analysis of the statistic is cumbersome. Therefore a statistic similar to CAV is formulated using the elements of $\Q$, the Bartlett decomposed SCM , where $\Rcap=\Q\Q^H$. 

\begin{equation}
 T_1(\Rcap)=\sum\limits_{1\leq j< i\leq N_r} |q_{ij}| \left/{\sum\limits_{k=1}^{N_r} q_{kk}}\right . \label{mod_QCAV}
\end{equation}
The Bartlett decomposition makes the elements of lower triangular $\Q$ matrix independent. They have the following distributional property under $\mathscr{H}_0$ \cite{mulvar_TWAnderson},
\begin{IEEEeqnarray}{rCl}
\frac{\sqrt{N}}{\sigma}q_{ii} & \sim  & \chi_{N-i+1},\quad \,\,\, 1\le i \le N_r\nonumber\\
\frac{\sqrt{N}}{\sigma}q_{ij} & \sim  &\mathcal{CN}(0,1),\quad  1\le j<i \le N_r\nonumber
\end{IEEEeqnarray}
where $\chi_k$ denotes a chi-random variable with $k$ degrees of freedom (d.o.f).
The independency between the numerator and denominator terms makes the analysis of the statistics formulated using the elements of $\Q$ simpler compared to the ones which use the dependent elements of $\Rcap$.
%
%
%
%

This motivates us to look into more possibilities of forming statistics with the elements of $\Q$.
The analysis presented in this paper considers non-parametric statistics on complex data which are designed as functions of elements of $\Q$ in the form of ratios similar to CAV and their combination. We show that the combined statistics are robust against uncertainties in the value of noise variance and correlation. Moreover, a few scenarios were identified under which these statistics exploit the correlation property better than the blind GLRT statistics and the CAV statistic resulting in improved performance. 

Section \ref{Prob_form} presents problem formulation, followed by analysis and observations about a number of non-parametric statistics in Section \ref{Non_para_test_stat}. 
Based on the analysis, we propose in section \ref{comb_stat_sec} combining these statistics leading to  improved performance and less sensitivity to variation in system parameters.
\section{Problem Formulation}\label{Prob_form}
Observation $\X\in \C^{N_r\times N}$ represents a block of $N$ samples across $N_r$ sensors, giving rise to the two hypotheses model: 
\begin{equation}
\X = \left\{
\begin{array}{ll}
\mathbf{Hs}+\boldsymbol{\eta}  \,,& \text{Signal hypothesis }  \mathscr{H}_1 \\
\boldsymbol{\eta} \,,& \text{Null hypothesis }\mathscr{H}_0
\end{array} \right.
\label{model_eq}
\end{equation}
where $\boldsymbol{\eta}{\in}\C^{N_r\times N}$ is the additive noise, which is assumed to be zero-mean circular complex Gaussian, spatially uncorrelated and temporally white with the covariance matrix $\sigma^2\I_{N_r}$. The signal transmitted from the $N_t$ number of sources (assuming $N_t{\le}N_r$) is temporally uncorrelated and  assumed to be i.i.d standard complex Gaussian vector, i.e, $\mathbf{s}\in\mathit{\C}^{N_t\times N}{\sim}\CN(0,\I)$. The channel present between the source and the sensor is assumed constant over the observation duration and modeled as correlated complex Gaussian matrix, i.e., $\H\in\C^{N_r\times N_t}$ wherein each $h_{ij}\sim \CN(0,1/N_t)$. The channel matrix $\H$ is modeled to capture the correlation that might be present between sensors, and one that is introduced by the channel. The correlation present in the model is assumed unknown. The SCM is calculated as $\Rcap{=}\X\X^H/N$.


Our aim is to design a statistic for the classification without having any information about the system parameters. We
consider non-parametric statistics shown in table \ref{4stat} whose formulations are similar to the CAV statistic in \eqref{mod_CAV}. They are called as Type 1, 2, 3 and 4 and are functions of $\Q$.

\begin{table}
\renewcommand{\arraystretch}{2.4}
\centering
\renewcommand{\tabcolsep}{0.5cm}
\caption{Statistics: Type 1, 2, 3 and 4}
  \begin{tabular}{|c|c|}
  \hline
\fbox{Type 1} & \fbox{Type 2}\\ 
$T_1=\dfrac{\sum\limits_{1\leq j< i\leq N_r} |q_{ij}| }{\sum\limits_{k=1}^{N_r} q_{kk}}$
&  
$T_2=\dfrac{\bigg| \sum\limits_{1\leq j< i\leq N_r}q_{ij} \bigg|}{\sum\limits_{k=1}^{N_r}  q_{kk} }$
\\ \hline
\fbox{Type 3} & \fbox{Type 4}\\ 
 $T_3=\dfrac{\left(\sum\limits_{1\leq j< i\leq N_r} | q_{ij}|\right)^2}{\sum\limits_{k=1}^{N_r}  q_{kk}^2 }$
&
$T_4=\dfrac{\bigg|\sum\limits_{1\leq j< i\leq N_r}q_{ij} \bigg|^2}{\sum\limits_{k=1}^{N_r}  q_{kk}^2 }$\\ 
  \hline
  \end{tabular}
\label{4stat}
  \end{table}

\section{Analysis of test statistics: Type 1, 2, 3 and 4}\label{Non_para_test_stat}
The exact closed form expression for the distribution of the statistic under the null hypothesis is required to find the detection threshold. Approximations are used when exact closed form expressions are not available and are intractable.
\subsection{Threshold Calculations}\label{threshold_calc}
\subsubsection{Type 1}
For the  $T_1$ statistic defined in Table \ref{4stat}, note that,   
$|q_{ij}| \sim$ Rayleigh $(1/\sqrt{2})$. 
The distribution of sum of these $N_R{=}N_r(N_r-1)/2$ independent Rayleigh random variables can be calculated as in \cite{ray_approx}. However, we approximate the sum distribution by the Gaussian tail approximation with the following parameters.
\begin{equation}
\mu_N =\frac{N_R}{\sqrt{2}}\sqrt{\frac{\pi}{2}},\quad
\sigma_N^2 =\left(2-\frac{\pi}{2}\right)\frac{N_R}{2} \label{norm_aprox}
\end{equation}
Since straightforward simulation show that the Gaussian tail approximation is more accurate than \cite{ray_approx} for the above case, the  threshold calculated with this approximation method is  used to evaluate $P_d$ in Fig.\ref{fig1}.
The denominator is trace of $\Q$, which is approximated to its mean value $\mu_\chi$ (see Appendix \ref{meantrace}) under moderately large sample size assumption.
Therefore, the parameters of the Gaussian distribution in \eqref{norm_aprox} are scaled by $1/\mu_\chi$ and  the threshold for $T_1$ for a given $\Pfa$ is given by, 
\begin{equation*}
\gamth{=}(\mu_N/\mu_\chi)+(\sigma_N^2/\mu_\chi^2) Q^{-1}(\Pfa) 
\end{equation*}
$Q$ denotes the tail probability of a Gaussian distribution.
\subsubsection{Type 2}
Using similar arguments,  $T_2$ can be shown to be a scaled 
 Rayleigh  with parameter $\sqrt{N_R/2}/\mu_\chi$. 
The Rayleigh CDF with this parameter is used to calculate the threshold.
\subsubsection{Type 3}
Note that the numerator of  $T_3$ is the square of the numerator of $T_1$ with approximate distribution given in \eqref{norm_aprox}. After squaring and scaling with $1/\sigma_N^2$,  it transforms to non-central $\chi^2$ with $1$ d.o.f and non-centrality parameter $\delta{=}(\mu_N/\sigma_N)^2$. The denominator is $\chi^2$ with $d=(N+1)N_r-N_r(N_r+1)/2$ d.o.f. 
When scaled properly the ratio follows $F_{1,{d}}(\delta)$ distribution, i.e.,
\begin{equation}
\frac{d}{\sigma_N^2} \,\,T_3\sim \frac{\chi^2_{1}(\delta)}{\chi^2_{d}/d}\sim F_{{1},{d}}(\delta)
\label{Type3_F}
\end{equation}
Similar statistic is proposed in \cite{blindet} for the real system model.
\subsubsection{Type 4}
When  the numerator of $T_4$ is scaled with $(N_r(N_r-1))^{-1}$, it follows $\chi^2$ distribution with $2$ d.o.f. The denominator is similar to \eqref{Type3_F}. Hence, 
\begin{equation}
\frac{d}{2\,N_r(N_r-1)}\,\,T_4\sim \frac{\chi^2_{2}/2}{\chi^2_{d}/d}\sim F_{{2},{d}}
\label{Type4_F}
\end{equation}

\subsection{Simulation set-up and Observations}\label{sim_obs}
\begin{figure*}
\centering
\subfigure[]
{
\includegraphics[width=.98\columnwidth, height=.62\columnwidth]{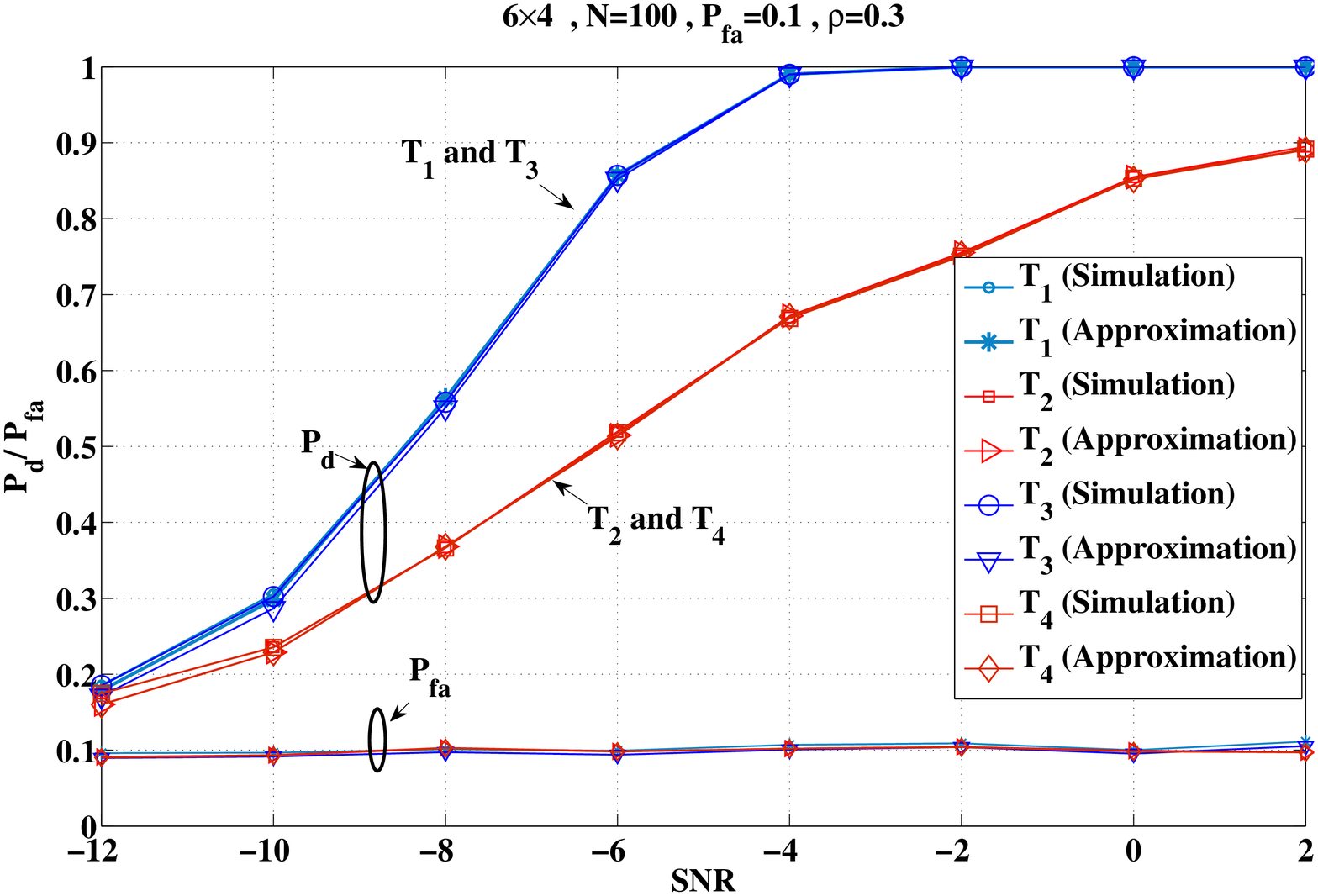}
\label{subfig1_11}
}
\hfil
\subfigure[]
{
\includegraphics[width=.98\columnwidth, height=.62\columnwidth]{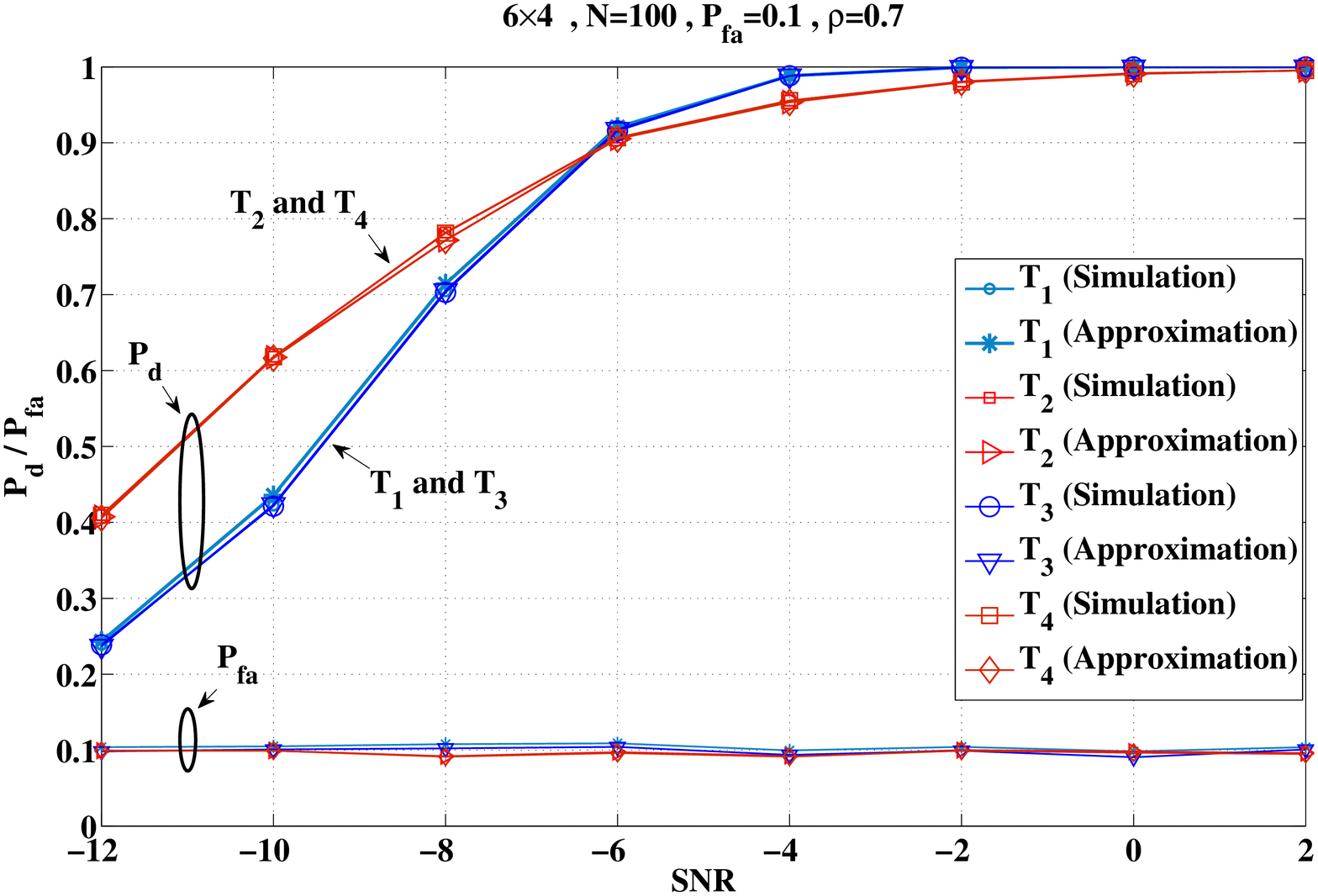}
\label{subfig1_21}
}
 \caption{$P_d$ vs. $SNR$ for \subref{subfig1_11} $\rho=0.3$ and \subref{subfig1_21} $\rho=0.7$}\label{fig1}
\end{figure*}

We consider $N_r{=}6$ and $N_t{=}4$. \cite{GLRT_Zhang} and \cite{mulAnt_abbas} among others assume large $N$ ($N\gg100$), we however assume moderate $N$ ($N \approx 100$) similar to \cite{rankP_ramirez} and \cite{Nadler_Report}. 
Since we consider the blind detection problem we assume that the structure of the covariance matrix under the signal hypothesis is not known at the detector. \emph{If the detector were not blind and knows the covariance structure, then the test statistic could be formed to exploit it}. We restrict our focus to the blind detection and impose AR(1) spatial covariance structure \cite{oestges2005impact} on the multichannel signal  for the simulation purpose (for comparison of different statistic’s ability to exploit correlation) such that,
\begin{equation*}
 \X=\P^{1/2}\mathbf{\tilde{H}s}+\boldsymbol{\eta}
\end{equation*}
$\P^{1/2}\tilde{\H}$ corresponds to $\H$ in \eqref{model_eq}, where elements of $\tilde{\H}$ are independent and $\tilde{h}_{ij}\sim \CN(0,1/N_t)$. $\P$ captures the correlation present in the system  and $\P^{1/2}$ is its Cholesky decomposition.
\begin{equation}
 \P =\begin{bmatrix}
1 & {\rho}     & {\rho^2} & {\ldots} \\
\rho     & 1 & {\rho}   & {\ldots} \\
\rho^2   &  \rho & 1    & {\ldots }\\
\vdots   & \vdots   & \vdots & \ddots
\end{bmatrix}\label{rho_mat}
\text{and}\quad \H=\P^{1/2}\mathbf{\tilde{H}}
\end{equation} 
The average received signal to noise ratio across each sensor is given by,
\begin{equation*}
 \text{SNR}=\frac{E[\tr(\mathbf{vv}^H)]}{E[\tr(\boldsymbol{\eta\eta}^H)]}\,\,,\text{ where} \,\,\mathbf{v}=\P^{\frac{1}{2}}\tilde{\H}\mathbf{s}
\end{equation*}
For each SNR, the distribution of the statistics $T_1,\,T_2,\,T_3$ and $T_4$ under the two hypotheses are obtained through 1000 Monte-Carlo realizations. The detection threshold is found using the null distribution of the statistics keeping a fixed constraint on the value of $\Pfa$ (=$0.1$). 
The performance ($P_d$) evaluated using this threshold is denoted as \textit{simulation} whereas $P_d$ obtained with the thresholds found in section \ref{threshold_calc} is denoted as \textit{Approximation} in Fig. \ref{fig1}. To verify the accuracy of approximations used in deriving the threshold, the $\Pfa$ is calculated using the null distribution of the statistic obtained through Monte-Carlo simulation and plotted along with $P_d$.

The calculations in section \ref{threshold_calc} indicate that the detection thresholds are independent of noise variance and Fig. \ref{fig1} verifies the accuracy of these calculations in maintaining $\Pfa$ at the preset value $0.1$. It also verifies the validity of threshold at sample size $N=100$.
In terms of performance we observe from Fig. \ref{fig1} that $T_1$ and $T_3$ are equivalent and $T_2$ and $T_4$ are equivalent. 
Note that both $T_2$ and $T_4$ have cross terms (refer \eqref{cross_term} in Appendix \ref{cor_coef_calc}) making them highly sensitive to correlation among streams. 

It would be desirable to have the performance independent of the correlation $\rho$ because correlation $\rho$ is not known. However, we desire to retain the good features of $T_1$ (or $T_3$) under low correlation and that of $T_2$ (or $T_4$) under high correlation. We now propose the combination statistics, $T_{12}=T_1+T_2$ and $T_{34}=T_3+T_4$ such that in $T_{12}$ (or $T_{34}$), $T_1$ (or $T_3$) will dominate under low correlation and $T_2$ (or $T_4$) will dominate under high correlation. 

\section{Combination of statistic}\label{comb_stat_sec}
\subsection{$T_{12}$ statistic}
When the denominator is replaced with the mean value $\mu_\chi$, the combination statistic effectively has mean and variance (from the similar arguments used in Type 1 and Type 2 threshold calculations) given as,
\begin{equation}
 \mu_{12} =\frac{N_R}{\sqrt{2}\mu_\chi}\sqrt{\frac{\pi}{2}}+\frac{\sqrt{N_R\,\pi}}{2\mu_\chi},\,\,
\sigma_{12} = \left(2-\frac{\pi}{2}\right)\frac{N_R}{\mu_\chi^2}\label{T12_diststat}
\end{equation}
The detection threshold using Gaussian tail approximation is,
\begin{equation*}
 \gamth=\mu_{12}+\sigma_{12} Q^{-1}(\Pfa)
\end{equation*}
 \subsection{$T_{34}$ Statistic}
From \eqref{Type3_F}, the scale $d/\sigma_N^2$ makes $T_3$ follow $F_{1,{d}}(\delta)$. 
Since the scaling should be same for both terms, $T_4$ after scaling is,
\begin{equation}
%
 \frac{d}{\sigma_N^2} \,T_4= \frac{\lvert\sum\limits_{1\leq j< i\leq N_r}q_{ij} \rvert^2/\sigma_N^2}{\sum\limits_{k=1}^{N_r}  q_{kk}^2 /d}\label{T34_frac}
\end{equation}
The numerator of \eqref{T34_frac} is Rayleigh random variable with parameter $\sqrt{N_R/2}/\sigma_N$. When squared,  it is distributed as exponential random variable with parameter $\Delta{=}2/(2-\pi/2)$ which is independent of system parameters $N_r$ and $N$. Therefore the ratio in \eqref{T34_frac} is scaled $F_{2,d}$ distribution with factor $\Delta$. 
Effectively the distribution of scaled $T_{34}$ is written as sum of two correlated F distributions (central and non-central), i.e., 
\begin{equation}
\frac{d}{\sigma_N^2}\,\,T_{34}\sim F_{1,d}(\delta)+\Delta\,F_{2,d} \equiv A+B\label{T34_AB}
\end{equation}
where $\equiv$ denotes termwise equivalence.
 The approximate correlation between the two terms is given in Appendix \ref{cor_coef_calc}.
If ($\mu_{A},\sigma^2_{A}$) and ($\mu_{B},\sigma^2_{B}$) are mean and variance of $A$ and $B$, the $\gamth$ can be calculated using Gaussian tail approximation, i.e.,
\begin{equation}
 \gamth=\mu_{34}+\sigma_{34} Q^{-1}(\Pfa)
\end{equation}
 where $\mu_{34}=\mu_{A}+\mu_{B}$ and $\sigma^2_{34}=\sigma^2_{A}+\sigma^2_{B}+2\rho_{AB}\,\sigma_{A}\sigma_{B}$.
%
%
\section{Simulation results and Discussion}\label{Sim_Res}
 We compare the performance of $T_{12}$ and $T_{34}$ with the CAV statistic and blind GLRT statistics, such as coherence ratio test and sphericity test. Also, we consider reduced sphericity test and RLRT which assume complete knowledge about the parameters $N_t$ and $\sigma^2$ respectively. This will enable us to know the loss in performance of the blind statistics for not knowing these parameters. Also, we analyse the sensitivity of the statistics to variation in system parameters at very low SNR ($-10$ dB).  The RLRT statistic is omitted in sensitivity comparison because it is less sensitive to variation in $\rho$ and $N_t$. The simulation set-up is similar to \ref{sim_obs} and chosen system parameters are indicated in each figure. 
\subsection{$P_d\text{ vs. }SNR$}
\begin{figure*}
\centering
\subfigure[]
{
\includegraphics[width=.98\columnwidth, height=.62\columnwidth]{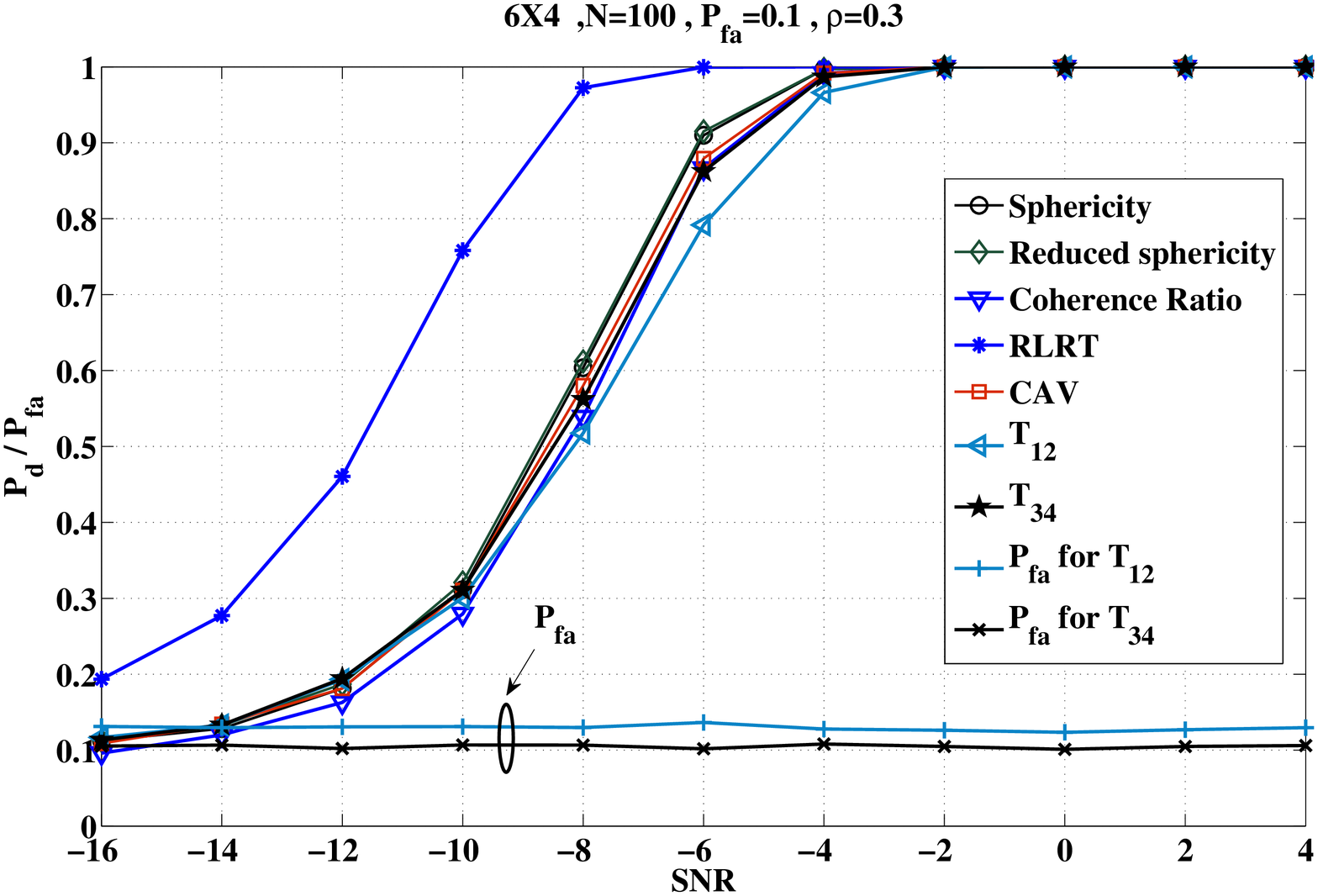}
\label{subfig2_11}
}
\hfil
\subfigure[]
{
\includegraphics[width=.98\columnwidth, height=.62\columnwidth]{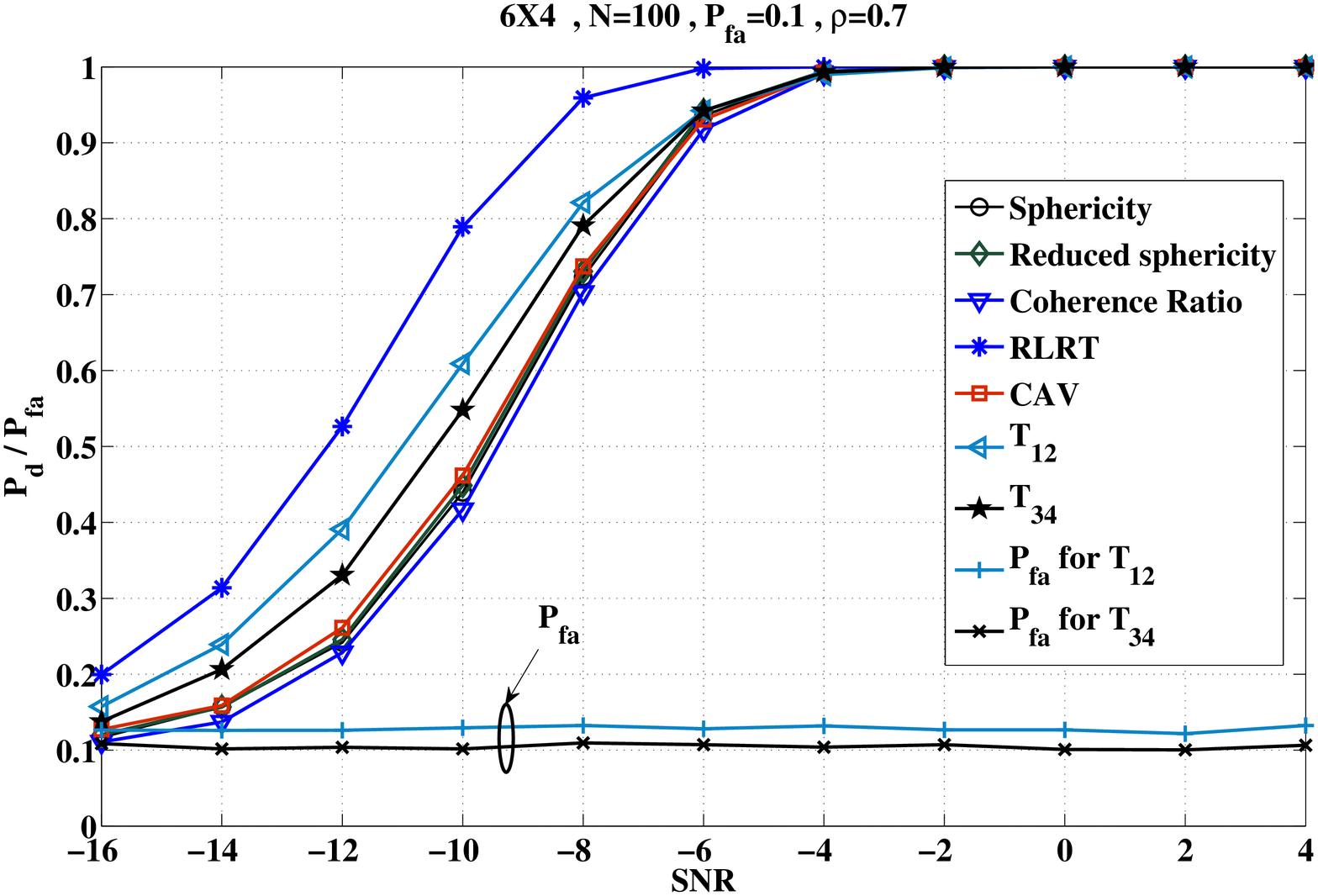}
\label{subfig2_21}
}
\caption{$P_d$ vs $SNR$ for \subref{subfig2_11} $\rho=0.3$ and \subref{subfig2_21} $\rho=0.7$}\label{fig2}
\end{figure*}

The performance of the statistics at different SNR under low correlation ($\rho=0.3$) and high correlation ($\rho=0.7$) is plotted in Fig. \ref{subfig2_11} and \ref{subfig2_21}. We observe that the performance of the $T_{34}$ statistic is equivalent to the  coherence ratio statistic (blind GLRT statistic) under low correlation and has a better performance compared to blind GLRT statistics under high correlation.  $T_{12}$ is advantageous compared to $T_{34}$ under high correlation, however, performs poorly under low correlation.
The complete knowledge about the noise variance makes the RLRT statistic perform better than blind statistics. The loss in performance due to lack of knowledge about the noise variance is significant under low correlation. Under high correlation the combination statistics reduce this loss by exploiting the spatial correlation. Moreover, 
the combination statistics exploit the correlation property better than the CAV statistic. 

The overshoot in $\Pfa$ indicates the effect of underestimation of the threshold. It is because the variance in \eqref{T12_diststat} is calculated neglecting the correlation between $T_1$ and $T_2$. The Gaussian tail approximation is accurate for the $T_{34}$ statistic.
It verifies the validity of threshold under low sample sizes ($N=100$) and also the robustness against the uncertainty in the value of noise variance. 
\subsection{Correlation among streams ($\rho$)}
\begin{figure}
\centering
{
\includegraphics[width=.98\columnwidth, height=.62\columnwidth]{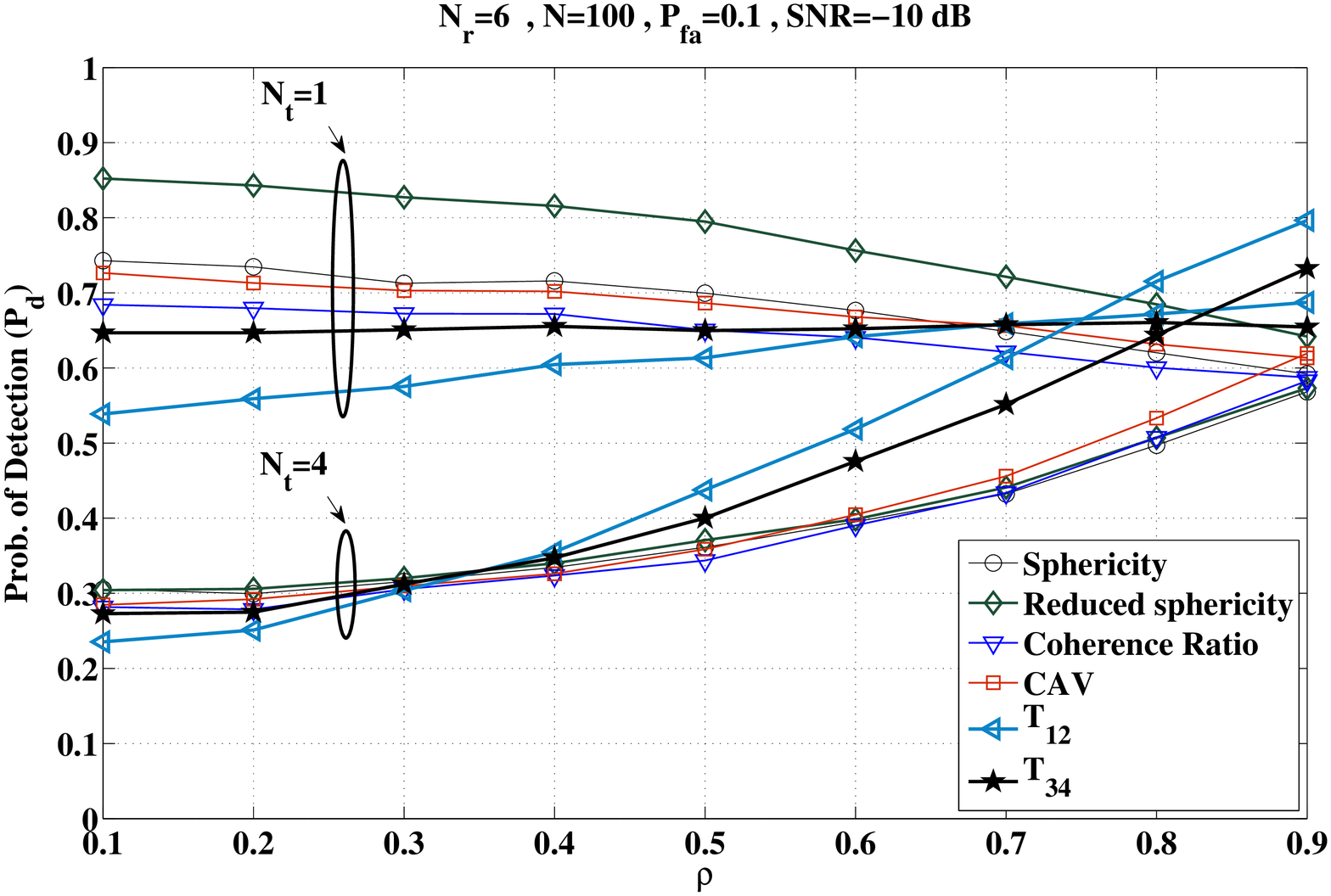}
}
\caption{$P_d$ vs. $\rho$ for $N_t=1$ and $N_t=4$}
\label{subfig2_1}
\end{figure}

It is expected that the detection performance should increase due to deviation in the observation's spherical structure as correlation among streams increases. The Fig. \ref{subfig2_1} depicts the effect of variation in the value of correlation on the performance of the statistics for a fixed number of sources in the system. If there are more than one source ($N_t{>}1$), increase in correlation improves the performance of  $T_{12}$ and $T_{34}$ statistic and this improvement is significant compared to the blind GLRT statistics. This shows that the combination statistics exploit the correlation property better than the blind GLRT statistics. However, when there exists only a single source ($N_t{=}1$, rank-1 channel), the correlation among channels in worse conditions results in decrease in performance ($P_d$) with increase in correlation $\rho$.  The combination statistics $T_{12}$ and $T_{34}$ perform poorly under this condition.  
\subsection{Number of sources ($N_t$)}
The performance of the statistics ($P_d$) decreases with increase in the number of sources \cite{Bayes_SS_CoDe}. This is due to the alignment of dominant right singular vectors of the channel in the statistical direction of the transmit covariance matrix, which is well known in MIMO literature by the name \emph{channel hardening effect} \cite{Ch_harden}. 

 The effect of variation in the number of sources on the performance of the statistics under low and high correlation is plotted in Fig. \ref{subfig4_11}. Under low correlation $T_{12}$ performs poorer than the blind GLRT statistics, however, it outperform all the other statistics (including $T_{34}$) under high correlation. The performance $T_{34}$ statistic is equivalent to blind GLRT statistics under low correlation and performs better than blind GLRT statistics under high correlation. The combination statistics are almost invariant to variation in $N_t$ under high correlation.   

\subsection{Sample size ($N$)}
\begin{figure*}
\centering
\subfigure[]
{
\includegraphics[width=.98\columnwidth, height=.6\columnwidth]{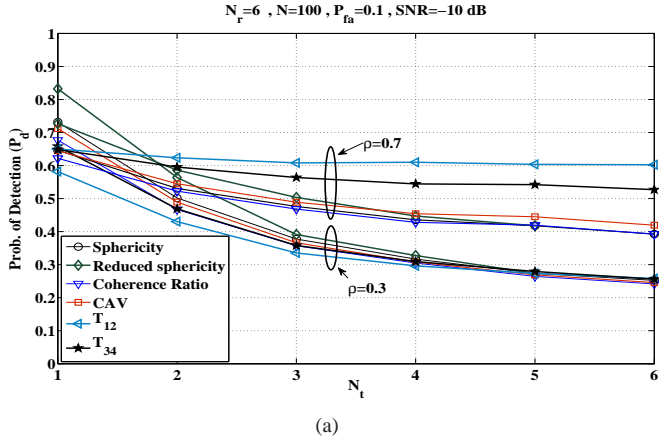}
\label{subfig4_11}
}
\hfil
\subfigure[]
{
\includegraphics[width=.98\columnwidth, height=.6\columnwidth]{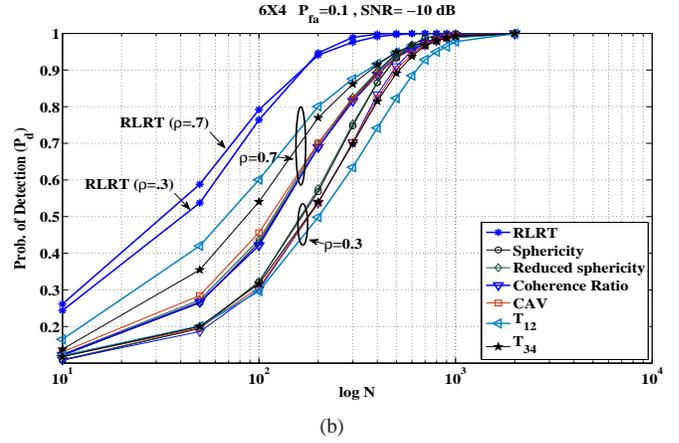}

\label{subfig4_21}
}
\caption{$P_d$ vs $N_t$ for $\rho=0.3$ and $0.7$, \subref{subfig4_21} $P_d$ vs $\log\, N$ for $\rho=0.3$ and $0.7$}\label{fig3}
\end{figure*}

The performance with variation in sample size ($N$) fixing the other two parameters $N_t$ and $\rho$ is plotted in  Fig. \ref{subfig4_21}. As expected, the detection probability for all the statistics approaches 1 as the sample size increases. The $T_{12}$ and $T_{34}$ statistics perform better than blind GLRT statistics under high correlation for all sample sizes. When low correlation scenario is considered $T_{12}$ performs poorer than GLRT statistics, however, $T_{34}$ statistic is equivalent to the blind GLRT statistics. Therefore, $T_{34}$ is the best choice if the statistic has to perform equally well under both high and low correlation scenario.  

\section{Conclusion}\label{conclusion}
The performance improvement for the considered multichannel detection problem, compared to sensitive and asymptotically optimal GLRT statistics, is achieved through combining the non-parametric statistics. The threshold calculations verifies the independent nature of the detection thresholds on the value of noise variance making the statistics robust to uncertainty in them. The Monte-Carlo simulation verifies it and also validates the approximation techniques used.

Under high correlation the proposed combination statistics have better performance compared to blind GLRT statistics and the CAV statistic from which all the designed statistics are motivated. Also, they are insensitive to variation in $N_t$ and have better performance at low sample sizes. Under low correlation, the performance of $T_{34}$ is equivalent to blind GLRT statistics, however, performance of $T_{12}$ is poorer than blind GLRT statistics. Therefore, if the statistic has to be chosen independent of correlation, $T_{34}$ would be a better choice. The only scenario where the combination statistics fail is when there exists only a single source ($N_t{=}1$) in the system. In such scenario, both $T_{12}$ and $T_{34}$ perform worse compared to blind GLRT statistics and CAV statistic. Extending the analysis to more general correlation model opens up many possibilities for future work.
%
%
%
\appendix
\subsection{Approximation for the trace of $\Q$}\label{meantrace}
When $k$ is large, the mean of $\chi_k$ random variable is,
\begin{equation*}
\mu=\sqrt{2}\,\,\dfrac{\Gamma\left((k+1)/2\right)}{\Gamma(k/2)}\approx \sqrt{k}\left(1-\frac{1}{4k}\right) 
\end{equation*}
The variance of $\chi_k$ random variable, ($k-\mu^2$) is far less compared to its mean $\mu$ when $k$ is moderately large. This is shown here.
\begin{equation*}
 \frac{k-\mu^2}{\mu}\approx\frac{k-k(1-\frac{1}{4k})^2}{\mu}\approx \frac{1}{2\mu}\ll 1
\end{equation*}
Hence, the trace of $\Q$ is replaced with its mean given by,
\begin{equation}
 \mu_{\chi}=E\left(\sum\limits_{k=1}^{N_r}  q_{kk}\right)=\sqrt{2}\,\,\sum_{i=1}^{N_r}\dfrac{\Gamma(\frac{N-i+2}{2})}{\Gamma(\frac{N-i+1}{2})}
\label{mu_chi}
\end{equation}
\subsection{Approximate correlation coefficient}\label{cor_coef_calc}
$q_{12},\,q_{13}\ldots\,q_{1 N_r},\,q_{23},\,q_{24}\ldots q_{N_r-1\, N_r}$ be represented as $a_1+jb_1,\,a_2+jb_2\ldots a_{N_R}+j\,b_{N_R}$. Note that they are all independent. 
Let $Y$ and $Z$ be numerator of $T_3$ and $T_4$ respectively, then
\begin{IEEEeqnarray}{rl}
Y & = \left(\sum\limits_{1\leq j< i\leq N_r} | q_{ij}|\right)^2\nonumber\\
& =\sum\limits_{k=1}^{N_R}(a_k^2+b_k^2)+2\sum\limits_{k\ne l}\,\sqrt{a_k^2+b_k^2}\sqrt{a_l^2+b_l^2}\nonumber\\
Z &=\bigg|\sum\limits_{1\leq j< i\leq N_r}q_{ij} \bigg|^2\nonumber\\
  &=\sum\limits_{k=1}^{N_R}(a_k^2+b_k^2)+\sum\limits_{k\ne l}\,\,(2a_k a_l+2b_k b_l)\label{cross_term}
\end{IEEEeqnarray}
The cross term is calculated as,
\begin{IEEEeqnarray}{rl}
 YZ = & \sum\limits_{k=1}^{N_R} (a_k^2+b_k^2)^2  +\sum\limits_{k\ne l} \,(a_k^2+b_k^2)\, (a_l^2+b_l^2)\nonumber\\
& +2\sum\limits_{k\ne l\ne m} \,(a_k^2+b_k^2)\, \sqrt{a_l^2+b_l^2} \sqrt{a_m^2+b_m^2}\nonumber\\
 & +2\sum\limits_{k\ne l} \,(a_k^2+b_k^2)^{3/2}\, (a_l^2+b_l^2)\nonumber\\
E[YZ] &=N_R[2+(N_R-1)(1+0.25\pi(N_R-2)+\sqrt{\pi}\,\Gamma{2.5})]\nonumber
\end{IEEEeqnarray}
The mean and variance of $Y$, $Z$ are calculated as,
\begin{align*}
\mu_Y &{=}N_R+0.25\pi N_R(N_R-1),\quad\sigma_Y^2=2+4\delta\\
\mu_Z &{=}N_R,\quad\sigma_Z^2=N_R^2
\end{align*}
where $\delta$ is the non-centrality parameter defined in \eqref{Type3_F}. $\rho_{YZ}$ is calculated using these parameters and $\rho_{AB}\approx\rho_{YZ}$.
\bibliographystyle{IEEEtran}
\bibliography{reference}
\end{document}